\documentclass[aps,prb,twocolumn,groupedaddress]{revtex4}
\usepackage[bookmarks]{hyperref}
\usepackage[dvips]{graphicx}
\usepackage{amsmath}
\usepackage{amstext}
\usepackage{graphics}
\usepackage{exscale}
\usepackage{epsfig}
\usepackage{changebar}
\usepackage[T1]{fontenc}
\usepackage{bm}
\usepackage{bbm}

\usepackage{version}

\begin{document}
\title[Renormalized perturbation theory for n-channel
Anderson model]{Renormalized parameters and perturbation theory for an n-channel
Anderson model with Hund's rule coupling: Asymmetric Case} 
\author{Y Nishikawa${}^{1,2}$ D J G Crow${}^1$ and A C Hewson${}^1$ }
\affiliation{${}^1$Department of Mathematics, Imperial College, London SW7 2AZ,
  UK.}
\affiliation{${}^2$Graduate School of Science, Osaka City University, Osaka 558-8585, Japan} 

\begin{abstract}

We explore the predictions of the renormalized perturbation theory for the
$n$-channel Anderson model, both with and without  Hund's rule coupling, in the regime away from
particle-hole symmetry. For the model with $n=2$ we deduce the renormalized
parameters from  numerical renormalization group calculations, and  plot them
as a function of the local occupation of the impurity site $n_d$. From these
we deduce the orbital, spin and charge susceptibilities, Wilson ratios and quasiparticle
density of states at $T=0$  in the
different parameter regimes, which gives  a
comprehensive overview of the low energy behavior of the model. We compare
the difference in Kondo behaviors at the points where $n_d=1$ and $n_d=2$. One
unexpected feature  of the  results is the suppression of the 
charge susceptibility in strong correlation regime over the occupation number
range $1\le n_d \le 3$.

\end{abstract}
\maketitle

\section{Introduction}

 In an earlier paper \cite{NCH10pre} we applied a renormalized perturbation approach 
to  study  the low temperature behavior
of an
$n$-channel impurity Anderson model with   a Hund's rule
exchange term. This form of perturbation theory is expressed in terms
of renormalized parameters of the model which have to be determined \cite{Hew93}.
We calculated these parameters explicitly for the particle-hole symmetric model with $n=2$
from numerical renormalization  group (NRG) calculations \cite{Wil75}. Here we extend that
work to calculate the renormalizations of the parameters away from
particle-hole symmetry. This enables us to compare the behavior of the model
in  regimes corresponding to different values of  occupation
number  at the impurity site, $n_d$ . These calculations reveal some unexpected
features in the variation of the renormalizations with $n_d$. For example,
when the
Hund's rule coupling $J_{\rm H}=0$ we find that  the points of maximum renormalization 
do not coincide with integral values of $n_d$ except at half-filling.
We also find  a strong suppression of the charge fluctuations when the
on-site
interaction is strong in regimes which would be classified as intermediate
or mixed valent. On substituting  these renormalized parameters into formulae
derived from the renormalized
perturbation theory (RPT), we can deduce the spin, orbital and charge
susceptibilities, specific heat coefficient and Wilson ratios at $T=0$ over the
full range of the occupation number $n_d$. This gives  a
comprehensive picture of the low energy behavior of the model, both with and
without the Hund's rule exchange term.\par
 We begin with a brief description of the model, and some of the 
results from the earlier work \cite{NCH10pre}, which will be used here. References
to this earlier paper will from here onwards be denoted by I. 
  The Hamiltonian takes the form,  
\begin{eqnarray}
{\cal H}=&&\sum_{m\sigma}\epsilon_{dm\sigma}d^{\dagger}_{m\sigma}d^{}_{m\sigma}+\sum_{k,m\sigma}\epsilon_{km\sigma}
c^{\dagger}_{k m\sigma} c^{}_{k m\sigma}\nonumber \\
&&+\sum_{k m\sigma} (V_k d^{\dagger}_{m\sigma} c^{}_{k m\sigma}
+  V_k^* c^{\dagger}_{k m\sigma} d^{}_{ m\sigma})+{\cal H}_d
\label{model1a}
\end{eqnarray}
where $d^{\dagger}_{ m\sigma}$, $d^{}_{ m\sigma}$, are creation and
annihilation operators for an electron in an impurity state  with total angular momentum
quantum number 
$l$, and $z$-component $m=-l,-l+1,  ... l$, where $2l+1=n$, the number of
channels,  and spin
component
$\sigma=\uparrow,\downarrow$.  
The creation and annihilation operators $c^{\dagger}_{km\sigma}$, $c^{}_{km\sigma}$ are
for  partial wave conduction electrons with energy
$\epsilon_{km\sigma}$. The hybridization width is determined by the 
factor  $\Delta_{m\sigma}(\epsilon)=\pi\sum_k|V_k|^2\delta(\epsilon-\epsilon_{km\sigma})$,
which we can take to be a constant $\Delta$ in the wide flat band limit.
The remaining part of the Hamiltonian,  ${\cal H}_d$,  describes the interaction between
the electrons in the impurity state, which we take to be of the form, 
\begin{eqnarray}
{\cal H}_d=&&
    {(U-J_{\rm H})\over 2}\sum_{mm'\sigma\sigma'} d^{\dagger}_{m\sigma}
    d^{\dagger}_{m'\sigma'} d^{}_{ m'\sigma'} d^{}_{ m\sigma}\nonumber  \\
&&+{J_{\rm H}\over 2}\sum_{mm'\sigma \sigma'} d^{\dagger}_{ m\sigma}
d^{\dagger}_{m'\sigma'} d^{}_{m\sigma'} d^{}_{ m'\sigma}. 
\label{model1b}
\end{eqnarray}
As well as the direct Coulomb interaction $U$ between the electrons,  a Hund's rule
exchange term $J_H$ is included between electrons in states with different $m$ values.
The sign for the exchange term has been chosen so that $J_H>0$ corresponds
to a ferromagnetic interaction.\par

 For the two-channel case ${\cal H}_d$ 
can be expressed in the form,
\begin{equation}
{\cal H}_d=U\sum_{\alpha=1,2}n_{d\alpha\uparrow}n_{d\alpha\downarrow}
   +U_{12}\sum_{\sigma\sigma'}n_{d,1\sigma}n_{d,2\sigma'}
-{2J_{\rm H}} {\bf S}_{d,1}\cdot{\bf S}_{d,2},  
\label{model2c}
\end{equation}
with a  ferromagnetic Heisenberg exchange coupling $2J_{\rm H}$ between the
electrons in the different channels, and $U_{12}=U-3J_{\rm H}/2$.\par 

The renormalized perturbation theory is formulated in terms of the
renormalized values of the parameters, $\epsilon_d$, $\Delta$, $U$,
and $J_{\rm H}$, which specify the model. We denote these by 
 $\tilde\epsilon_d$, $\tilde\Delta$, $\tilde U$,
and $\tilde J_{\rm H}$. They are defined in terms of the self-energy of the
impurity
Green's function and the 4-vertices at zero frequency. We will not
repeat the definitions here but refer to I.
 The impurity specific heat coefficient $\gamma$, the spin $\chi_s$,
orbital $\chi_{\rm orb}$,
and charge  $\chi_c$ susceptibilities at $T=0$ (zero magnetic field) can all be expressed explicitly
in terms of these parameters.\par
The specific heat coefficient $\gamma$ is given by
\begin{equation}
 \gamma=2n\pi^2\tilde \rho^{(0)}(0)/3.
\end{equation} 
where $\tilde\rho^{(0)}(\omega)$ is the free 
quasiparticle density of states per single spin and channel,
\begin{equation}
\tilde \rho_{m\sigma}^{(0)}(\omega)={\tilde\Delta/\pi\over (\omega-\tilde\epsilon_{d})^2 +\tilde\Delta^2}.\label{fqpdos}
\end{equation}
 The results for the spin
susceptibility  is given by 
\begin{equation}
\chi_s=2n\mu_{\rm B}^2\eta_s\tilde \rho^{(0)}(0),
\label{chis}
\end{equation}
where 
\begin{equation}
\eta_s=1+(\tilde U+(n-1)\tilde J_{\rm H})\tilde\rho^{(0)}(0).
\label{chiseta}
\end{equation}
Similarly for the orbital susceptibility,
\begin{equation}
\chi_{orb}={(n^2-1)\mu_{\rm B}^2\eta_{orb}\tilde \rho^{(0)}(0)\over 12},
\label{chiorb}
\end{equation}
where
\begin{equation}
\eta_{orb}=1+(\tilde U-3\tilde J_{\rm H})\tilde\rho^{(0)}(0),
\label{chiorbeta}
\end{equation}
and the charge susceptibility,
\begin{equation}
\chi_c=2n\eta_c\tilde \rho^{(0)}(0),
\label{chic}
\end{equation}
where
\begin{equation}
\eta_c=1-((2n-1)\tilde U-3(n-1)\tilde J_{\rm H})\tilde\rho^{(0)}(0).
\label{chiceta}
\end{equation}

The total occupation of the impurity site  $n_d$ at $T=0$ is given by 
\begin{equation}
 n_{d}={2}-{4\over \pi}{\rm tan}^{-1}\left
({\tilde\epsilon_d\over \tilde\Delta}\right),
\label{qpocc}
\end{equation}
which corresponds to the Friedel sum rule.
Using the result in equation (\ref{qpocc}), we can derive an expression for
$\rho^{(0)}(0)$ in terms of the total occupation of the impurity site,
$n_d$, 
\begin{equation}
\tilde\rho^{(0)}(0)={{\rm sin}^2({\pi n_d/2n})\over \pi\tilde\Delta}.
\label{qpdosnd}
\end{equation}
These  results can all be shown to be exact for the model with the
renormalized parameters as defined in I.\par
\section{NRG calculation of parameters for n=2}
\begin{figure}[!htbp]
   \begin{center}
     \includegraphics[width=0.34\textwidth]{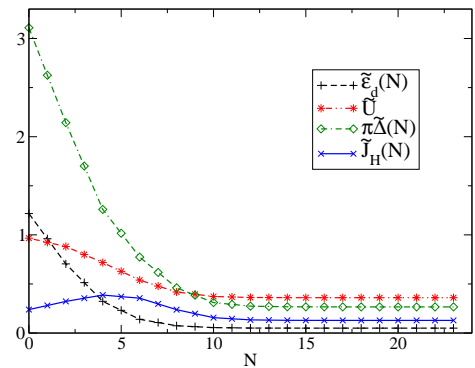}
     \caption{(Color online) A plot of  $\tilde \epsilon_d(N)$,  $
       \pi\tilde\Delta(N)$, $\tilde U(N)$ and  $\tilde J_{\rm H}(N)$ (in units of
   $\pi\Delta$) as a function of $N$ the NRG iteration number
   for  $U/\pi\Delta=4$, $J_{\rm H}/\pi\Delta=0.15$, $\pi\Delta=0.01$
and $\epsilon_d/\pi\Delta=-3.574$. The renormalized parameters $\tilde\epsilon_d$,  $
       \pi\tilde\Delta$, $\tilde U$ and  $\tilde J_{\rm H}$ are given by the
   fixed point values corresponding to the plateau region which develops for large  $N$.
} 
     \label{vsN}
   \end{center}
 \end{figure}

To evaluate the  formulae  for the low temperature properties of the model 
we need the values for the renormalized parameters. As shown in I
an accurate way of calculating these in terms of the bare parameters
$\epsilon_d$, $\Delta$, $U$ and $J_{\rm H}$, is from an analysis of  
the approach to the low energy fixed point of the Wilson
numerical renormalization group calculation. This method  can be applied for channel
numbers $n=1,2$, but  becomes progressively more difficult to impossible
for
larger values of $n$, due to the increase in the size of the matrices to be
diagonalized.
If the renormalized parameters are defined as a function of $N$, the number of
NRG iteration steps,
then the renormalized values correspond to the fixed point values for large
$N$.  We use this approach for the $n=2$ model, as we did for the
particle-hole symmetric case. In that case we could take $\tilde\epsilon_d=0$,
but here, in moving away from particle-hole symmetry, we  
need to determine the additional parameter $\tilde\epsilon_d$.
In Fig. \ref{vsN} we show a typical plot for the case  $U/\pi\Delta=4$,
$J_{\rm H}/\pi\Delta=0.15$, $\pi\Delta=0.01$ and $\epsilon_d/\pi\Delta=-3.574$. The
renormalized parameters can be deduced accurately from the plateau regions
that develop for large $N$. For more details we refer to I and the references therein.

\subsection{SU(2$n$) Model $J_{\rm H}=0$}

We first of all look at the results for the model with 
 $J_{\rm H}=0$, which has $SU(2n)$ symmetry. Our main interest will be in 
the strong correlation regime, where $U$ is large and the impurity electrons
are almost localized, such that the charge susceptibility is suppressed.
If we  take $\eta_c=0$
in Eq. (\ref{chiceta})
we find  
\begin{equation}
\tilde U\tilde\rho^{(0)}(0)={1\over (2n-1)}.
\label{relationa}
\end{equation}
This implies that the effect of the quasiparticle interactions gets weaker
the larger the channel index $n$, and goes to zero in the limit $n\to\infty$.
Substituting  the expression for the quasiparticle density of states given in Eq.
 (\ref{fqpdos}) into Eq. (\ref{relationa}) gives
 \begin{equation}
\tilde U={\pi\tilde\Delta\over (2n-1){\rm sin}^2({\pi n_d/ 2n})
 }.
\label{ratio1}
\end{equation}
If Eq. (\ref{relationa})  is satisfied then  $\eta_s$ from Eq. (\ref{chiseta})
is equal to $2n/(2n-1)$. As $\eta_s$ coincides with the definition of the
Wilson ratio, $R_{\rm W}=\pi^2\chi_s/3\mu_{\rm
  B}^2\gamma$, we find   $R_{\rm W}=2n/(2n-1)$
when we are in a localized regime.\par
\begin{figure}[!htbp]
   \begin{center}
     \includegraphics[width=0.34\textwidth]{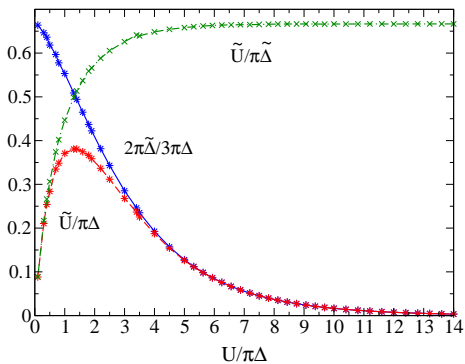}
     \caption{(Color online) A plot of  $\tilde U/\pi\Delta$,  $2\tilde
       \Delta/3\Delta$ and  $\tilde U/\pi\tilde\Delta$
the case $n_d=1$ with $U/\pi\Delta$, $J_{\rm H}=0$ and  $\pi\Delta=0.01$} 
     \label{nd1u}
   \end{center}
 \end{figure}

For $n=2$ we expect these relations to be satisfied at the points of integral valence
$n_d=1,2,3$. In I we have already shown from the NRG
calculations for $n=2$ that at the particle-hole symmetric point $n_d=2$,
the relation $\tilde U/\pi\tilde\Delta=1/3$ in agreement with
Eq. (\ref{ratio1}). For $n_d=1,3$, from Eq.  (\ref{ratio1})
we get the result $\tilde U/\pi\tilde\Delta=2/3$. 
 In
 Fig. \ref{nd1u}  we plot $2\tilde\Delta/3\Delta$, $\tilde U/\pi\Delta$,
and the ratio  $\tilde U/\pi\tilde\Delta$
for $n_d=1$ as a function of  $ U/\pi\Delta$ for $\pi\Delta=0.01$.
It can be seen that there is localization and  a single energy scale for $ U/\pi\Delta>4.5$,
and the ratio  $\tilde U/\pi\tilde\Delta$ asymptotically approaches the
value $2/3$. For   $ U/\pi\Delta=14$ we find $\tilde U/\pi\tilde\Delta= 0.66665
$, a very accurate verification of the relation from Eq. (\ref{ratio1})
for $n_d=1$ and $n=2$.
\par
\begin{figure}[!htbp]
   \begin{center}
     \includegraphics[width=0.42\textwidth]{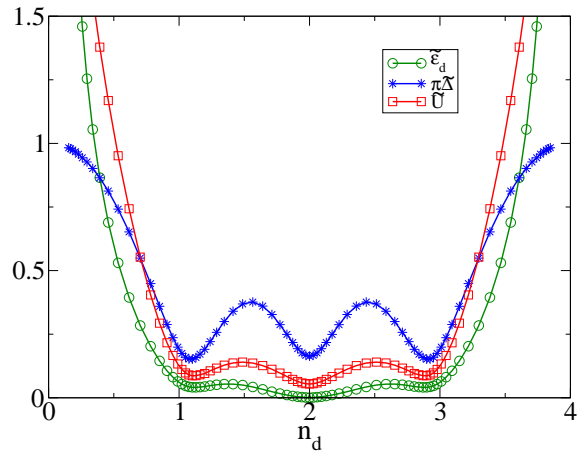}
     \caption{(Color online)  The renormalized parameters
       $\tilde\epsilon_d$, $\pi\tilde\Delta$, 
$\tilde U$  (in units of $\pi\Delta$ with $\Delta=0.01$)
as a function of the impurity occupation $n_d$ for 
  $U/\pi\Delta=5$ and  $J_{\rm H}=0$.
} 
     \label{rdata5}
   \end{center}
 \end{figure}

To look at the behavior more generally in the strong correlation regime,
we have calculated the renormalized parameters for $U/\pi\Delta=5$,
$\Delta=0.01$ over the full range of the occupation number $n_d$.
The results for $\tilde\epsilon_d/\pi\Delta$,
$\tilde \Delta/\Delta$ and $\tilde U/\pi\Delta$   are shown in Fig. \ref{rdata5}. There  are three distinct local minina in
$\tilde\Delta$ at $n_d=2$ and for values of $n_d$ slightly greater than 1 and
slightly less than 3. This was to be expected, as the regions near integral values of
$n_d$ for large $U$ correspond to localized Kondo regimes,
and the dips in the values of $\tilde \Delta$ indicate  a  narrowing of
the quasiparticle density of states at these points. It is an unexpected
result, however, that minima away from the particle-hole symmetric point
$n_d=2$ 
are not precisely at $n_d=1$ and $3$.
Also we find in the mixed valence
regimes, $1<n_d<2$ and $2<n_d<3$, there is still
some significant renormalization of $\Delta$. For $n_d < 0.7$
and $n_d > 3.3$, the values of $\tilde\Delta$ rapidly approach the bare value
$\Delta$. \par

\vspace*{1.9cm}

\begin{figure}[!htbp]
   \begin{center}
     \includegraphics[width=0.36\textwidth]{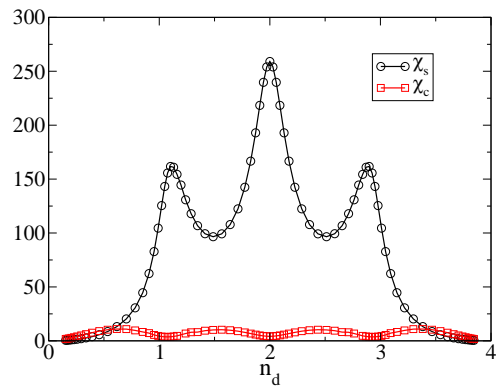}
     \caption{(Color online) The spin susceptibility $\chi_s$ (units of
       $4\mu_{\rm B}^2$) and the charge susceptibility
       $\chi_c$
as a function of the impurity occupation $n_d$ for 
  $U/\pi\Delta=5$ and  $J_{\rm H}=0$.
} 
     \label{chi5}
   \end{center}
 \end{figure}

In Fig. \ref{chi5} we give the corresponding results for $\chi_s$ and 
$\chi_c$. The enhanced peaks in $\chi_s$  in the Kondo regimes near integral
values of $n_d$ are as expected. It can be seen from Fig. \ref{rdata5} that the values of
$\tilde\Delta$ at $n_d=1,2,3$ are almost the same.  The higher value of
$\chi_s$ at $n_d=2$, therefore, is due to the
fact that at this point $\tilde\epsilon_d=0$  giving a higher 
quasiparticle density of states  compared with the peaks near  $n_d=1$
and  $n_d=3$. Again we note that  the peaks in $\chi_s$
near $n_d=1$ and $n_d=3$ are not precisely at these integer values.  \par

 \noindent

 \begin{figure}[!htbp]
   \begin{center}
     \includegraphics[width=0.36\textwidth]{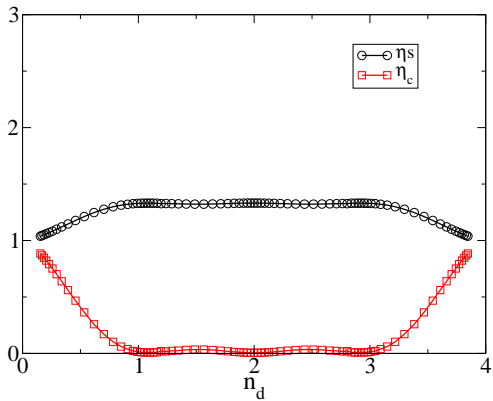}
     \caption{(Color online)  The coefficients $\eta_s$ ($=R_{\rm W}$, Wilson
       ratio), and $\eta_{c}$
as a function of the impurity occupation $n_d$ for 
  $U/\pi\Delta=5$ and  $J_{\rm H}=0$.
} 
     \label{eta5}
   \end{center}
 \end{figure}
 \noindent
 
The values of $\chi_c$ can be seen to be very small  at $n_d=2$ and
near  $n_d=1,3$. It is, however, rather small over the {\em whole range},
with only  modest peaks between the integer values of $n_d$.
To get more insight into this, we plot $\eta_s$ and $\eta_c$ for the same set of
parameters 
in Fig. \ref{eta5}. Surprisingly  we see that $\eta_c$ is very small over the whole range
from $n_d=1$ to $n_d=3$, and not just at the integer values. This indicates
that quasiparticle interaction suppresses the charge susceptibility 
not just for $n_d=1,2,3$ but also in the mixed valence regimes
 $1< n_d<2$ and  $2< n_d<3$. There are slight peaks in $\eta_c$
near $n_d\sim 1.5$ and  $n_d\sim 2.5$, but the values are still very small.
The Wilson ratio for this model when $n=2$ is equal to  $4/3$. It can be seen that $R_{\rm W}\approx 4/3$
 over the range $1\le n_d \le 3$. The fact the $\eta_c$
is very small  implies that impurity d-electrons are rather localized so that
Eq. (\ref{ratio1}) should be  a good approximation  over this range.
 For $n=2$, Eq. (\ref{ratio1}) becomes
\begin{equation}
{\tilde U\over\pi\tilde\Delta}={1\over 3{\rm sin}^2(\pi n_d/4)}.
\label{ratio2}
\end{equation}

 \begin{figure}[!htbp]
   \begin{center}
     \includegraphics[width=0.34\textwidth]{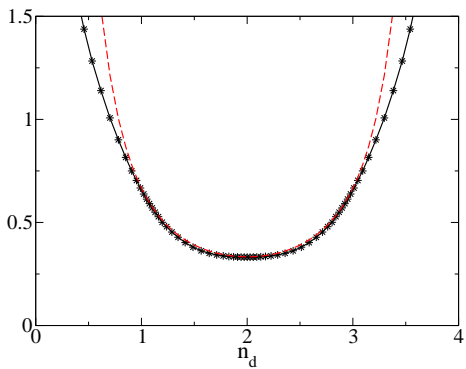}
     \caption{(Color online) A comparison of $\tilde U/\pi\tilde\Delta$ 
  versus $n_d$ for  $U/\pi\Delta=5$,  $J_{\rm H}=0$ and  $\Delta=0.01$
 (full curve with stars) with that derived using the  formula  in 
 Eq. (\ref{ratio2})  (dashed curve). }
     \label{ratio5}
   \end{center}
 \end{figure}
 \noindent
In Fig. \ref{ratio5}
we plot $\tilde U/\pi\tilde\Delta$  and compare with the form
given on the right-hand side of Eq. (\ref{ratio2}). It can be seen that the two
curves
are in good agreement over the range from $n_d=0.9$ to $n_d=3.1$.\par

It is also of interest to compare the two integral valence cases, $n_d=1$ 
and  $n_d=2$. 
In the Kondo regime for large $U$, the models with $n_d=1$ and $n_d=2$
can both be mapped onto  Coqblin-Schrieffer models of the form,
\begin{equation}
{\cal H}_{\rm CS}= J_{\rm eff} \sum_{\nu,\nu',k,k'}Y_{\nu,\nu'}c^{\dagger}_{k',\nu'}c^{}_{k,\nu}
+ \sum_{\nu,k}\epsilon_{k}c^{\dagger}_{k,\nu}c^{}_{k,\nu},
\label{Kondo_model}\end{equation}
where the sum over $\nu=1,2, ... 2n$, and  with particle -hole symmetry
 $J_{\rm eff}=
4|V|^2/U$. The operators $Y_{\nu,\nu'}$ obey
the SU(2n) commutation relations,
\begin{equation}
  [Y_{\nu,\nu'}, Y_{\nu'',\nu'''}]_-=Y_{\nu,\nu'''}\delta_{\nu',\nu''}-Y_{\nu'',\nu'}\delta_{\nu,\nu'''},
\end{equation}
with  $\sum_\nu Y_{\nu,\nu}=nI$. 
  Though we are dealing with  the model for the same value of $n$, in this
  case  $n=2$, the models for
  $n_d=1$ and $n_d=2$, differ in that the operators
transform according to different irreducible representations of $SU(4)$.
 The case with  $n_d=1$ is the one originally considered by Coqblin and
Schrieffer \cite{CS69}, where the representation of the  operators
$Y_{\nu,\nu'}$ is the fundamental representation of the group,
which for the SU(4) group has dimension 4. On the other hand 
for $n_d=2$, as we noted in I, the  $Y_{\nu,\nu'}$ operators
 correspond to a 6 dimensional irreducible
representation of the SU(4) group.
This is similar to a Heisenberg model, which can describe
physical situations depending on the dimensionality of the irreducible
representation used for the spin operators, $2S+1$ for a spin $S$. 
  In the general $n$ channel model  with $r$ localized electrons
the dimensionality of the irreducible representation of the operators in the
Coqblin-Schrieffer model  will be $(2n)!/(2n-r)!r!$. \par
More generally when $n_d\le 1$, and $U/\pi\Delta\gg 1$, the model with $J_{\rm
  H}=0$ can be related to the $N$-fold degenerate, $U=\infty$,
Anderson model which has been applied to rare earth impurities
such as Ce and Yb with $N=2n$ \cite{OTW83,Sch89,Hew93b}. In this application the index $\nu$
corresponds the $z$-component  of total angular momentum, orbital plus spin $m_j$,
with $N=2j+1$, and $j$ is the total angular momentum quantum number.
The equation for the total angular momentum susceptibility
$\chi_j$ for this model is  given by
\begin{equation}
\chi_j={(g\mu_{\rm B})^2j(j+1)\over 3}N\eta_j\tilde \rho^{(0)}(0),
\label{chij}
\end{equation}
where $g$ is the $g$-factor for coupling to the magnetic field and  
\begin{equation}
\eta_j=1+\tilde U\tilde\rho^{(0)}(0),
\label{chijeta}
\end{equation}
while the equation for the charge susceptibility  is the
same as in Eq. (\ref{chic}) with $2n=N$. 
The Wilson ratio for the $N$-fold degenerate  model ($U=\infty$) is defined as 
$R_{\rm W}=\pi^2\chi_j/j(j+1)(g\mu_{\rm
  B})^2\gamma$, giving  $R_{\rm W} =\eta_j=N/(N-1)$, which is the same
as that we have for the SU(2n) model for $N=2n$.\par
In the localized limit when $U/\pi\Delta$ is large we have only one energy scale which we take
to be the Kondo temperature $T_{\rm K}$. For the $N$-fold degenerate, infinite
$U$, Anderson model  $T_{\rm K}$ is 
defined  such that  $\chi_j=(g\mu_{\rm
  B})^2j(j+1)/3T_{\rm k}$. 
This is equivalent to 
\begin{equation}
T_{\rm K}= {2n-1\over 4n^2 \tilde\rho^{(0)}(0)},
\label{TK1}
\end{equation}
which we will take as a general definition for $T_{\rm K}$ for the SU(2n)
model in the discussion here. It differs from the definition used
in I by the factor $(2n-1)/n^2$. \par
\begin{figure}[!htbp]
   \begin{center}
     \includegraphics[width=0.36\textwidth]{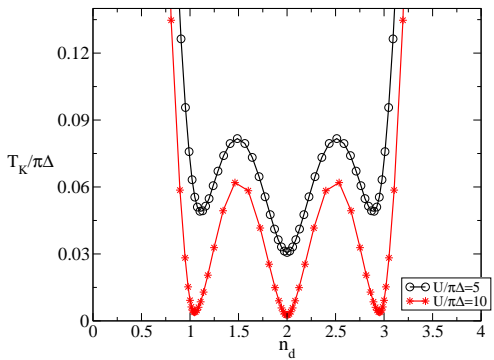}
     \caption{(Color online) A plot of the Kondo temperature $T_{\rm K}$,
as defined in Eq. (\ref{TK2}), as a function of $n_d$. The formula given in
(\ref{TK2})
is valid over the interval  $0.9< n_d<3.1$ where $\eta_c \approx 0$. }
     \label{TK5}
   \end{center}
 \end{figure}

 
On using Eq. (\ref{qpdosnd})  for $\tilde\rho^{(0)}(0)$, we find
\begin{equation}
T_{\rm K}={\pi\tilde\Delta(2n-1)\over 4 n^2{\rm sin}^2({\pi n_d/ 2n})
 }.\label{TK2}
\end{equation}
We would expect this formula to apply only at or near the points of integer
occupation of the impurity site, $n_d=1,2,3$. However, for large $U/\pi\Delta$
we found localization and a Wilson ratio $R_{\rm W}\approx 4/3$ over the complete
range $1\le n_d\le 3$. This means that we can define a Kondo temperature
as a function of $n_d$ over this range. A plot of $T_{\rm K}$ based on Eq. (\ref{TK2})
is given in Fig. \ref{TK5} for $n=2$ and $U/\pi\Delta =5,10$ ($\Delta=0.01$).
 There are three dips corresponding to a local minima for $T_{\rm K}$
at $n_d=2$ and near $n_d=1,3$. For the larger value of $U$ the outer minima
move slighter closer towards  $n_d=1$ and $n_d=3$.\par 

\begin{figure}[!htbp]
   \begin{center}
     \includegraphics[width=0.4\textwidth]{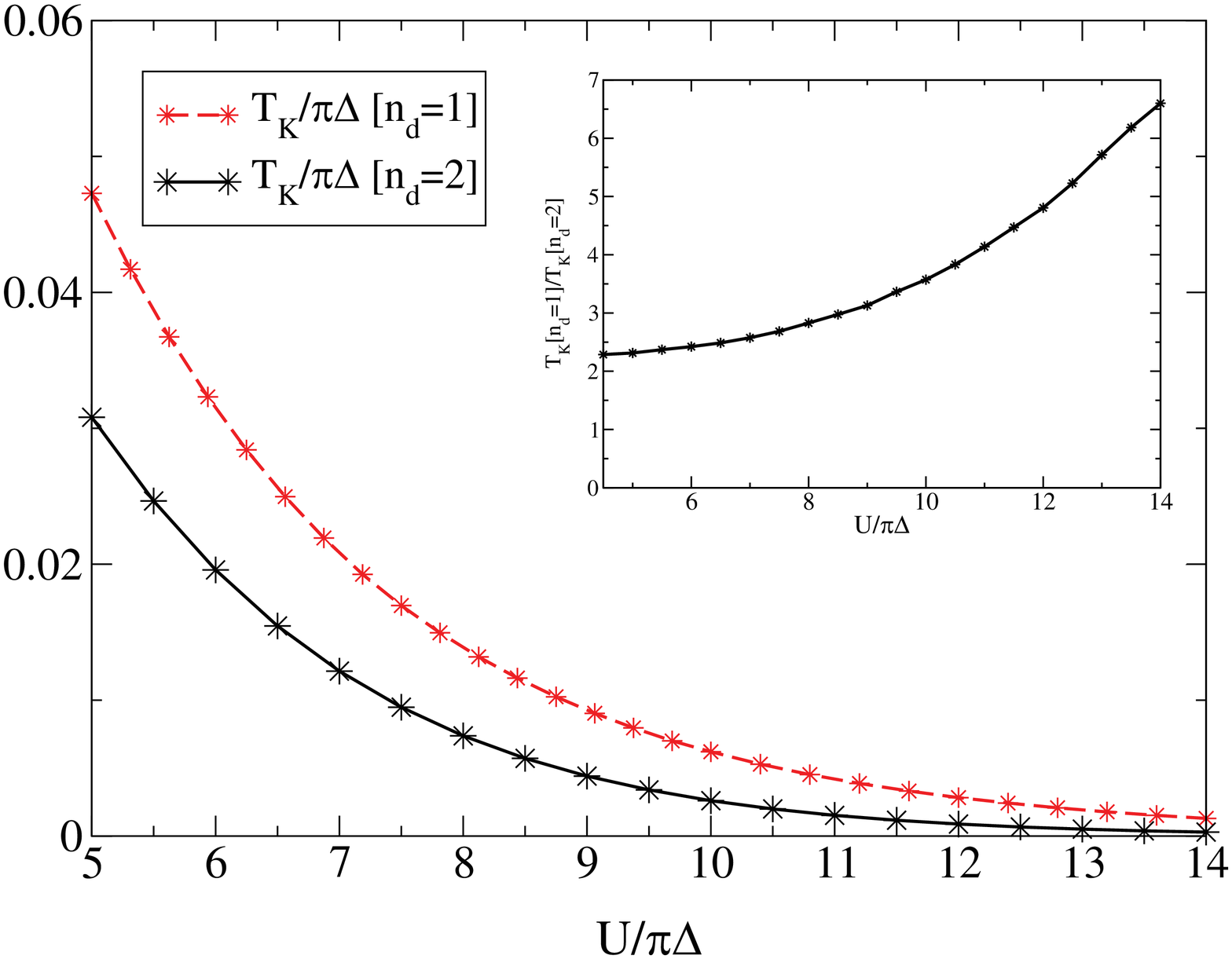}
     \caption{(Color online) The  Kondo temperatures $T_{\rm K}$
for   $n_d=1$ and 
   $n_d=2$  as a function of 
 for $U/\pi\Delta$ for  $J_{\rm H}=0$ and  $\pi\Delta=0.01$.
The inset shows the corresponding ratio of $T_{\rm K}[n_d=1]/T_{\rm K}[n_d=2]$
} 
     \label{tkratio}
   \end{center}
 \end{figure}

 \noindent
We noted earlier that in the Kondo limit $n_d=1$ and $n_d=2$ are
  described by different Coqblin-Schrieffer models. They  also have different values for the Kondo temperature $T_{\rm
  K}$. 
In Fig. \ref{tkratio} we  plot the Kondo temperatures $T_{\rm K}$ for
$n_d=1$ and $n_d=2$
 for the range  $5\le U/\pi\Delta\le
14$ for $\pi\Delta=0.01$.  In I we fitted the $T_{\rm K}$ for $n_d=2$
to the exponential form, $T_K/\pi\Delta={\rm const}\times ue^{-\pi^2u/16+0.25/u}$
where $u=U/\pi\Delta$.  The ratio of $T_{\rm K}$ for $n_d=1$ to that for
$n_d=2$ is shown in the inset of   Fig. \ref{tkratio}, and is seen to increase
monotonically with $U/\pi\Delta$.
\par
\vspace*{1.9cm}
 \noindent
In the Kondo regime for general $n$ we can deduce  the parameters $\tilde\epsilon_d$ and $\tilde U$ in terms of  $T_{\rm K}$,
\begin{equation}
\tilde \epsilon_d=T_{\rm K}{2n^2{\rm sin}(\pi n_d/n)\over \pi (2n-1) }.
\label{ee}
\end{equation}
\begin{equation}
\tilde U=\left({2n\over 2n-1}\right)^2T_{\rm K}.
\label{uu}
\end{equation}
Using  Eq. (\ref{ee}) and  (\ref{TK2})  we can derive an explicit expression  the quasiparticle density of states $\tilde\rho^{(0)}(\omega)$
in
the Kondo regime, 
\begin{equation}
\tilde\rho^{(0)}(\omega)={(2n-1)/ 4n^2T_{\rm K}\over
  (\Omega-{\rm cos}(\pi n_d/2n))^2+{\rm sin}^2(\pi n_d/2n) },
\label{rhosu2n}
\end{equation}
where $\Omega=\omega\pi(2n-1)/4T_{\rm K}n^2{\rm sin}(\pi n_d/2n)$.
In applying the results in Eqs. (\ref{TK2}) to (\ref{rhosu2n})  to the infinite $U$
model we must take $n_d=1$.\par

We can contrast the quasiparticle density of states in the case of
half-filling,
 $n_d=n$, with that for  $n_d=1$.  In the former case,
$\tilde\epsilon_d=0$ and the quasiparticle density of states is symmetrically
placed about the Fermi level, and for large $n$, $\tilde\rho^{(0)}(\omega)$ takes the approximate form,
\begin{equation}
\tilde\rho^{(0)}(\omega)\approx {2nT_{\rm K}/\pi^2\over
 \omega^2+(2nT_{\rm K}/\pi)^2 }.
\end{equation}
For $n_d=1$ and $n>1$, on the other hand, the quasiparticle peak is
asymmetrically placed about the Fermi level. For large $n$,  $\tilde\rho^{(0)}(\omega)$
takes the approximate form,
\begin{equation}
\tilde\rho^{(0)}(\omega)\approx {T_{\rm K}/2n\over
  (\omega-T_{\rm K})^2+(\pi T_{\rm K}/2n)^2 },
\end{equation}
and in the limit $n\to\infty$ collapses to a delta-function at a point
$T_{\rm K}$ above the Fermi level. This asymmetry with respect to the Fermi
level  for $n_d=1$, $n>1$, is required by the Friedel sum rule, because
if $n_d=1$, the quasiparticle density of states must be such that
 has only a fraction $1/2n$ is filled when integrated up  to the Fermi level.
\par

\vspace*{1.9cm}
 \begin{figure}[!htbp]
   \begin{center}
     \includegraphics[width=0.34\textwidth]{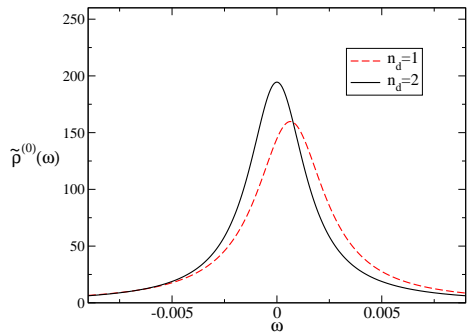}
     \caption{(Color online) A comparison of the quasiparticle density of
       states
$\tilde\rho^{(0)}(\omega)$  for  $n_d=2$ and $n_d=1$ for $U/\pi\Delta=5$, $J_{\rm H}=0$ and  $\Delta=0.01$} 
     \label{rho5}
   \end{center}
 \end{figure}
 \noindent

 In Fig. \ref{rho5} we compare
the 
 quasiparticle density of states for  $n_d=1$ and $2$ for  $U/\pi\Delta=5$.
We see that the peak in the case $n_d=1$ is shifted slightly above the Fermi
level so that the Friedel sum rule is satisfied in terms of the
quasiparticles.
This shift has physical consequences. If we ignore the effects of the
quasiparticle interactions, which becomes an increasingly good approximation the
larger the channel index $n$, we can estimate the low temperature and low
magnetic field
corrections to the spin susceptibility from a free quasiparticle calculation.
The $T^2$ and $H^2$ correction to the spin susceptibility, as well as the 
$T^3$ correction of the impurity specific heat contribution from this
calculation are proportional to  
\begin{equation}
\left({\rho^{(0)''}(0)\over
    \rho^{(0)}(0)}\right) -\left({\rho^{(0)'}(0)\over \rho^{(0)}(0)}\right)^2={2(\tilde\epsilon_d^2-\tilde\Delta^2)\over(\tilde\epsilon_d^2+\tilde\Delta^2)^2},\label{coeff}
\end{equation}
where $\rho^{(0)'}(0)$ and $\rho^{(0)''}(0)$ are the first and second derivatives of 
 $\rho^{(0)}(\omega)$ evaluated at $\omega=0$.
If $|\tilde\epsilon_d|<\tilde \Delta$, which is the case when
$\tilde\epsilon_d=0$, then this coefficient is negative. However, when the
quasiparticle density of states becomes asymmetric about the Fermi level such 
that $|\tilde\epsilon_d|>\tilde \Delta$ it changes sign to become positive.
As the susceptibility must eventually decrease at high temperatures and in
high magnetic fields, this implies that there must be a peak in $\chi_s(T)$
and $\chi_s(H)$. Such a peak in found in the exact Bethe ansatz solutions
for the $N$-fold degenerate, infinite $U$, Anderson model for $N>3$ \cite{OTW83,Raj83,Hew93b}, 
and this simple argument provides a qualitative explanation for this behavior.
 The form of the quasiparticle density of states in the vicinity of the Fermi
 level  also affects the thermopower. The thermopower due to the
impurity is proportional to the gradient of the quasiparticle density of states
at the Fermi level; it is zero when the quasiparticle density of states is
symmetrical about the Fermi level but large when there is narrow peak
just above the Fermi level. \par
It is also possible that shift in the peak in the
 quasiparticle density of states, which for $n=2$ will be to above the Fermi
 level for $n_d=1$ and below the Fermi level for $n_d=3$, may explain why the
local minima in  renormalized parameters and the peaks in the spin
 susceptibility do not occur at precisely at $n_d=1$ and $n_d=3$, but in one
 case slightly greater than $n_d=1$ and in the other slightly less than
 $n_d=3$.\par

  \subsection{Model with $J_{\rm H}\neq 0$}

\begin{figure}[!htbp]
   \begin{center}
     \includegraphics[width=0.42\textwidth]{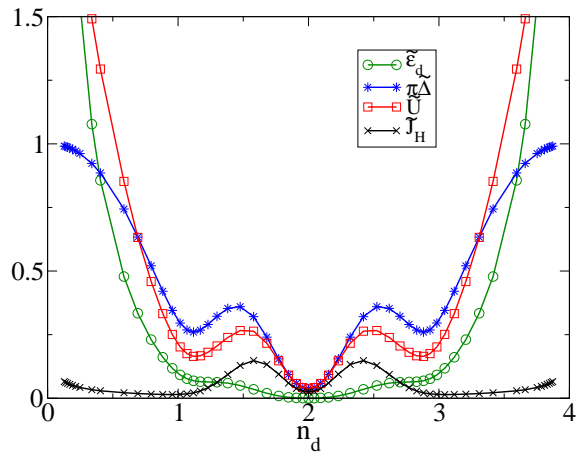}
     \caption{ (Color online) The renormalized parameters $\tilde\epsilon_d$,
 $\tilde\Delta$, $\tilde U$ and $\tilde J_{\rm H}$ (in units of $\pi\Delta=0.01$)
as a function of the impurity occupation $n_d$ for 
  $U/\pi\Delta=4$ and  $J_{\rm H}/\pi\Delta=0.15$. } 
     \label{rdata1}
   \end{center}
 \end{figure}

\begin{figure}[!htbp]
   \begin{center}
     \includegraphics[width=0.42\textwidth]{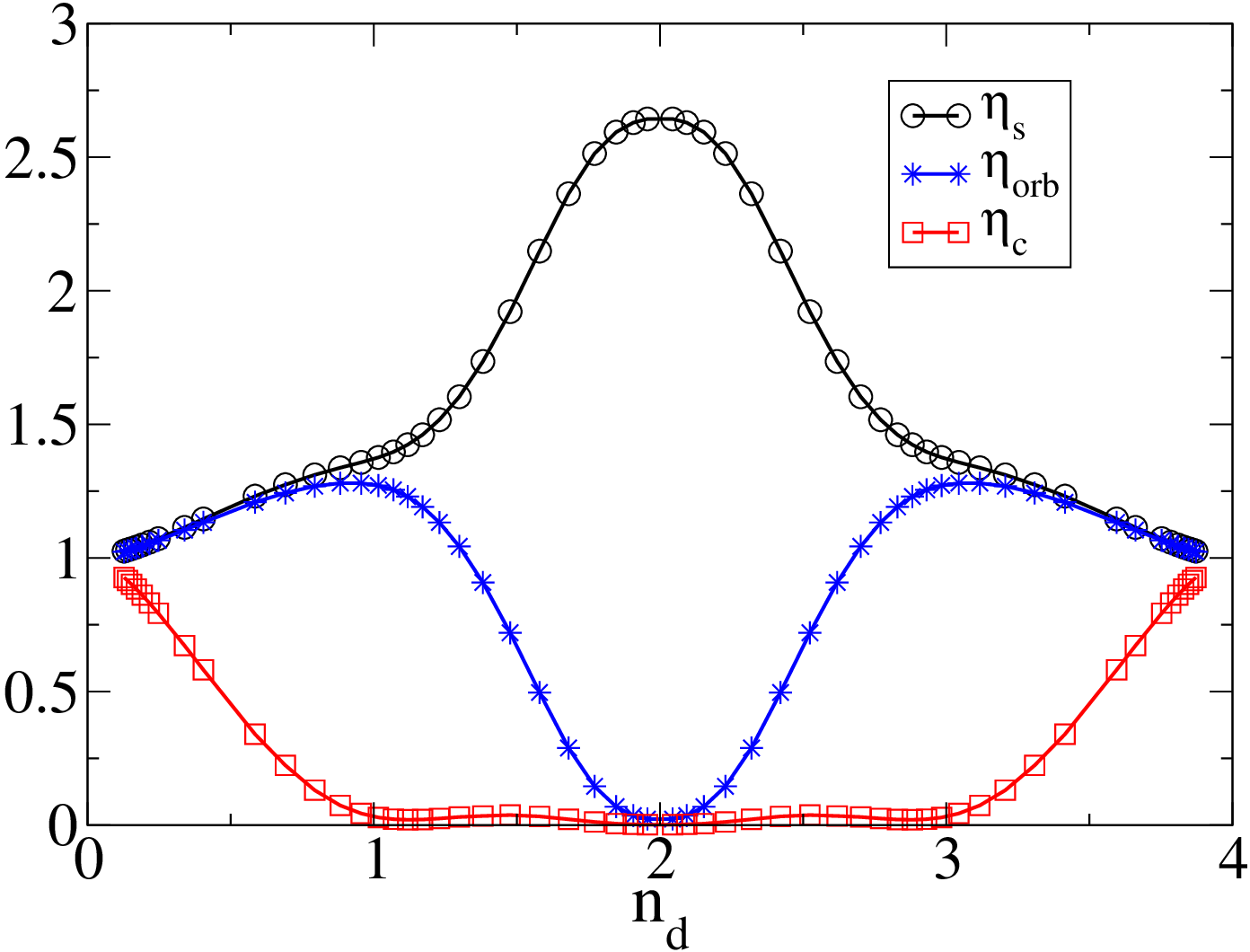}
     \caption{ (Color online) The coefficients $\eta_s$ ($=R_{\rm W}$, Wilson
       ratio),  $\eta_{orb}$, and $\eta_{c}$
as a function of the impurity occupation $n_d$ for 
  $U/\pi\Delta=4$ and  $J_{\rm H}/\pi\Delta=0.15$. } 
     \label{eta1}
   \end{center}
 \end{figure}

We now consider the case with Hund's rule coupling away from particle-hole
symmetry. In Fig. \ref{rdata1} we show the renormalized parameters $\tilde\epsilon_d$, $\tilde\Delta$,
$\tilde U$ and $\tilde J_{\rm H}$ (in units of $\pi\Delta=0.01$)
as a function of the impurity occupation $n_d$ for 
  $U/\pi\Delta=4$ and  $J_{\rm H}/\pi\Delta=0.15$. The main difference with
  those seen in Fig. \ref{rdata5} where $J_{\rm H}=0$ is that the minimum at $n_d=2$
is much more pronounced than the local minima near $n_d=1,3$. 
 The plot of
$\eta_s$ (=$R_{\rm W}$, Wilson ratio), $\eta_{orb}$  and $\eta_c$  in
Fig. \ref{eta1} shows
that the quasiparticle interaction from the Hund's rule coupling  induces an enhancement of $\eta_s$
and a corresponding reduction of  $\eta_{orb}$ in the range between
$n_d=1$ and $n_d=3$. Due to the relatively large value of $U/\pi\Delta$
the value of $\eta_c$ is almost completely suppressed between
$n_d=1$ to $n_d=3$ as in the case with $J_{\rm H}=0$.\par
As $\eta_c$ is very small over the range $1\le n_d\le 3$, we can to a good
approximation  equate it
 to zero.  From Eq. (\ref{chiceta}) this gives  a relation between the
 renormalized parameters, which for $n=2$
is
\begin{equation}
3(\tilde U-\tilde J_{\rm H})\tilde\rho^{(0)}(0)=1.
\label{relation3}
\end{equation}
In the discussion of the SU(2n) model, where $J_{\rm H}=0$, this condition left  only one independent renormalized parameter,
which we could take as the Kondo temperature. However, when $J_{\rm H}\ne 0$,
we are left with two renormalized parameters, so we cannot in this case 
define a Kondo temperature from this equation alone. We see from
Fig. \ref{eta1} that at and very close to $n_d=2$ the orbital susceptibility
is also suppressed, so equating $\eta_{\rm orb}$ to zero from
Eq. (\ref{chiorbeta}) we find
\begin{equation}
(3\tilde J_{\rm H}-\tilde U)\tilde\rho^{(0)}(0)=1.
\label{relation4}
\end{equation}
At this point we have a single energy scale and can define a Kondo temperature
via $\chi_s={(g\mu_{\rm B})^2S(S+1)/ 3T_{\rm K}}$ for a spin $S=1$, which
is such that
\begin{equation}
\pi\tilde\Delta=\tilde U={3\over 2}\tilde J_{\rm H}=4T_{\rm K},\label{relation2}
\end{equation}
and $\tilde\epsilon_d=0$ from particle-hole symmetry. At this point the Wilson
ratio $R_{\rm W}=\eta_s =8/3$, as can seen in Fig. \ref{eta1}. The
particle-hole symmetric case with $n_d=2$ is discussed more fully in I.
\par

\begin{figure}[!htbp]
   \begin{center}
     \includegraphics[width=0.42\textwidth]{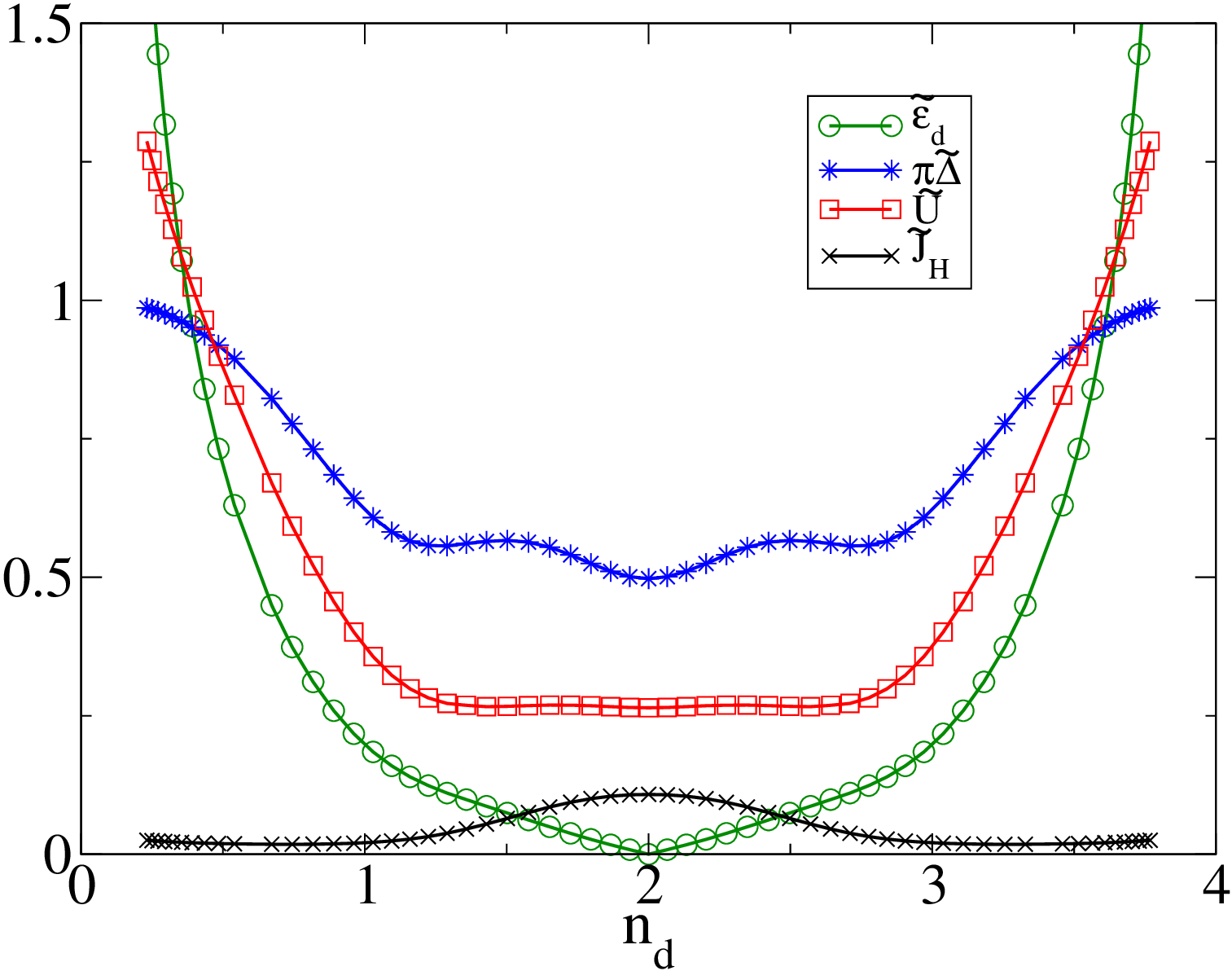}
     \caption{ (Color online) The renormalized parameters $\tilde\epsilon_d$,
 $\tilde\Delta$,
$\tilde U$ and $\tilde J_{\rm H}$ (in units of $\pi\Delta=0.01$)
as a function of the impurity occupation $n_d$ for 
  $U/\pi\Delta=2$ and  $J_{\rm H}/\pi\Delta=0.05$. } 
     \label{rdata2}
   \end{center}
 \end{figure}

\begin{figure}[!htbp]
   \begin{center}
     \includegraphics[width=0.42\textwidth]{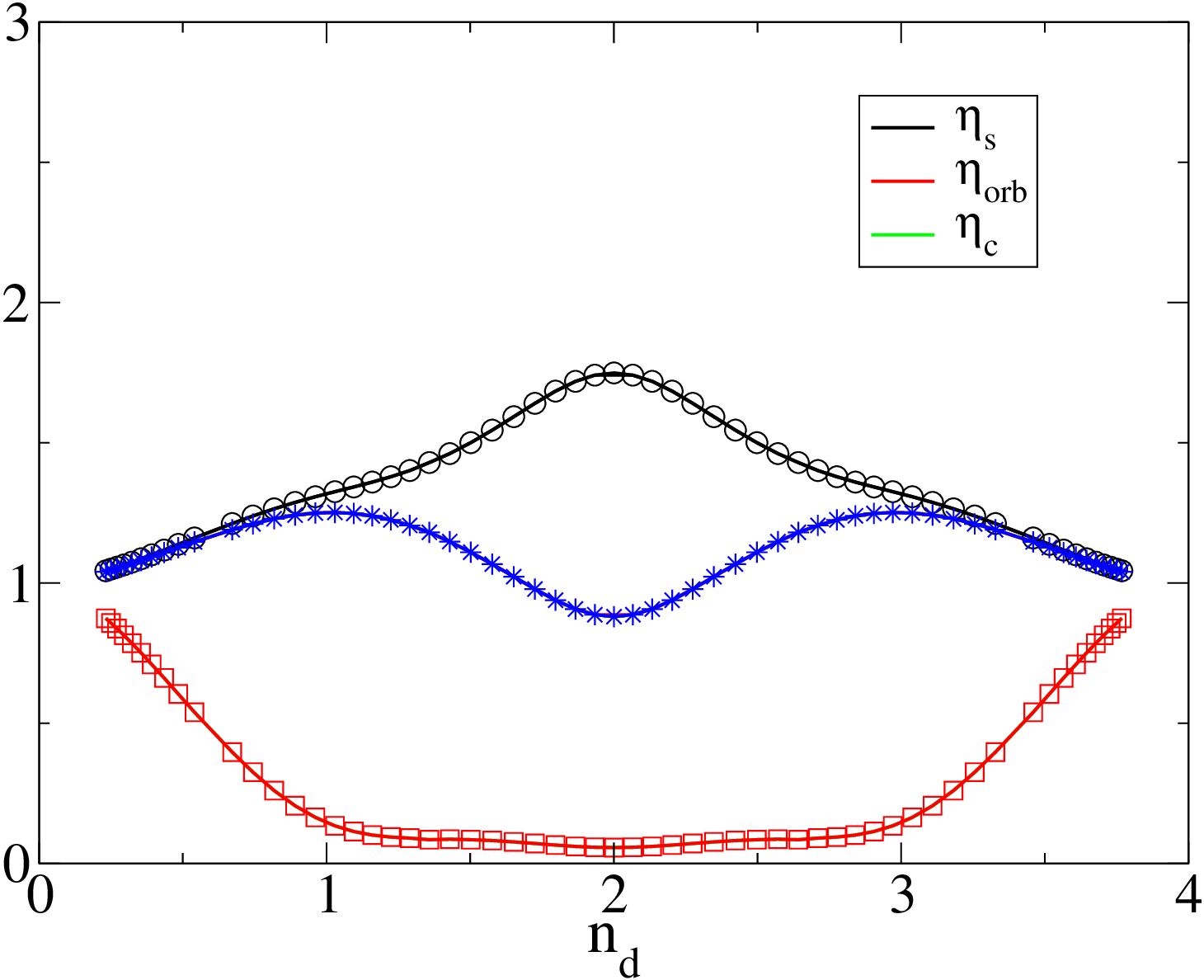}
     \caption{ (Color online) The coefficients $\eta_s$ ($=R_{\rm W}$, Wilson
       ratio),  $\eta_{orb}$, and $\eta_{c}$
as a function of the impurity occupation $n_d$ for 
  $U/\pi\Delta=2$ and  $J_{\rm H}/\pi\Delta=0.05$.
 } 
     \label{eta2}
   \end{center}
 \end{figure}

In Fig. \ref{rdata2} we show the renormalized parameters as a function of $n_d$  for smaller values of $U$
and $J_{\rm H}$, $U/\pi\Delta=2$ and  $J_{\rm H}/\pi\Delta=0.05$.
The values of  $\tilde\epsilon_d$, $\tilde\Delta$,
$\tilde U$ are rather similar to the case with $J_{\rm H}=0$ shown in
Fig. \ref{rdata5}. In this case, in contrast to the previous example, $\tilde
J_{\rm H}$ has a maximum at $n_d=2$ rather than a minimum. For this smaller
value of $U$, the renormalized value $\tilde U$ is rather flat in most of
the range from $n_d=1$ to $n_d=3$.
 In Fig. \ref{eta2}
we show the corresponding values for $\eta_s$, $\eta_{orb}$ and $\eta_c$.
It shows that,  even for  this smaller value of $U$, there is some suppression
of the charge susceptibility by the quasiparticle interactions. There is also some
enhancement of $\chi_s$ and a commensurate reduction in   $\eta_{orb}$ 
as the particle-hole symmetric point $n_d=2$ is approached.\par
\begin{figure}[!htbp]
   \begin{center}
     \includegraphics[width=0.42\textwidth]{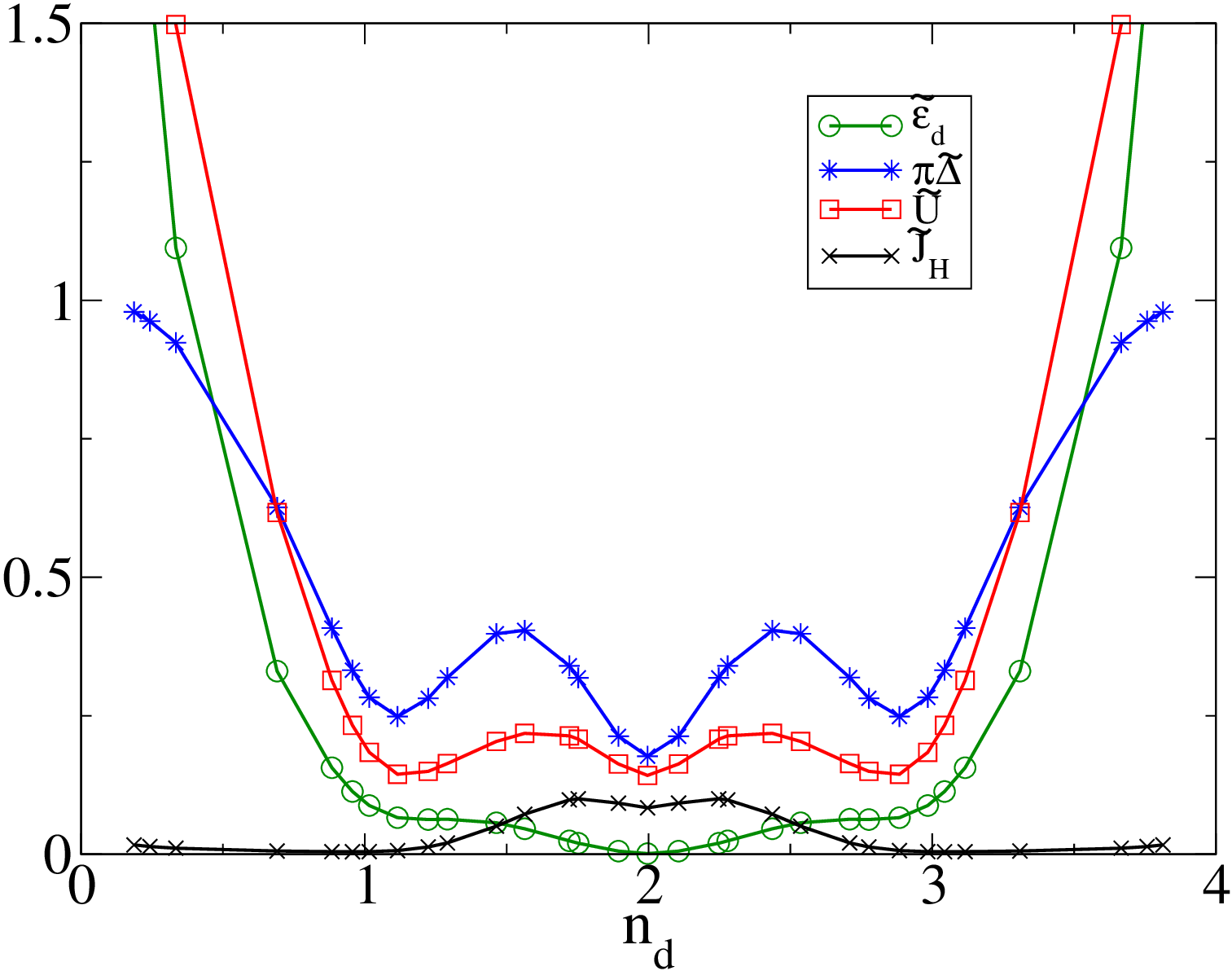}
     \caption{ (Color online) The renormalized parameters $\tilde\epsilon_d$,
 $\tilde\Delta$,
$\tilde U$ and $\tilde J_{\rm H}$ (in units of $\pi\Delta=0.01$)
as a function of the impurity occupation $n_d$ for 
  $U/\pi\Delta=4$ and  $J_{\rm H}/\pi\Delta=0.05$.   } 
     \label{rdata3}
   \end{center}
 \end{figure}
\begin{figure}[!htbp]
   \begin{center}
     \includegraphics[width=0.42\textwidth]{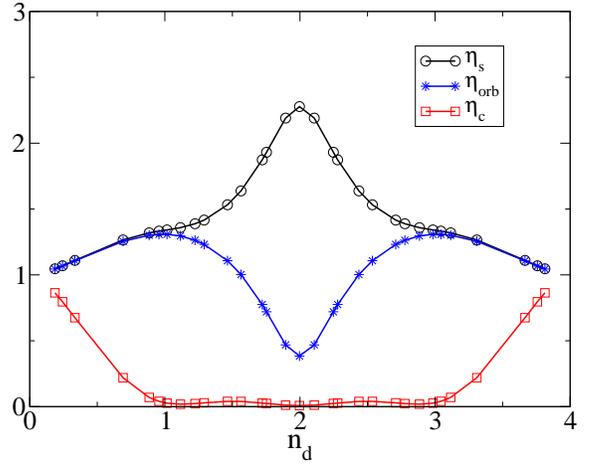}
     \caption{ (Color online) The coefficients $\eta_s$ ($=R_{\rm W}$, Wilson
       ratio),  $\eta_{orb}$, and $\eta_{c}$
as a function of the impurity occupation $n_d$ for 
  $U/\pi\Delta=4$ and  $J_{\rm H}/\pi\Delta=0.05$.
 } 
     \label{eta3}
   \end{center}
 \end{figure}

In Fig. \ref{rdata3} we show the renormalized parameters with the same value
of $J_{\rm H}$ ($J_{\rm H}/\pi\Delta=0.05$) and a larger value of $U$,
$U/\pi\Delta=4$. It can be seen that the effect of increasing $U$ is to
induce a rather shallow minimum in $\tilde J_{\rm H}$ at $n_d=2$,
and also in $\tilde U$.  In Fig. \ref{eta3}
we give the corresponding values for $\eta_s$, $\eta_{orb}$ and $\eta_c$.
It can be seen that, despite using the same value of $J_{\rm H}$, 
 $\eta_s$ is enhanced and  $\eta_{orb}$ is reduced in the central region
for the larger value of $U$. This indicates that the enhancement of $\eta_s$
does not scale in proportion to $J_{\rm H}/U$, but that $J_{\rm H}$ is more
effective in suppressing the orbital fluctuations when the charge fluctuations
are suppressed by a larger value of $U$.\par
\begin{figure}[!htbp]
   \begin{center}
     \includegraphics[width=0.42\textwidth]{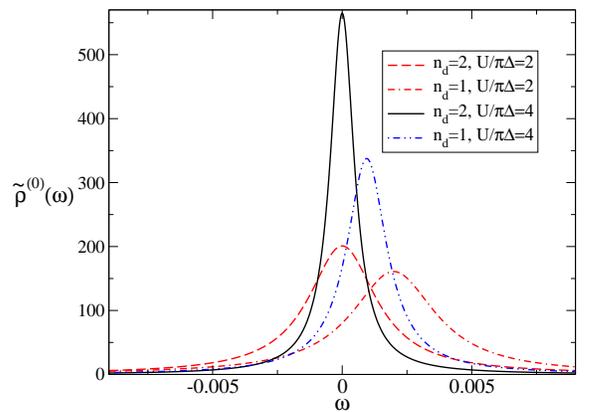}
     \caption{ (Color online) The quasiparticle density of states
       $\tilde\rho^{(0)}(\omega)$  for $J_{\rm H}/\pi\Delta=0.05$,
       $U/\pi\Delta=2,4$ and
       $n_d=1,2$
 } 
     \label{rho3}
   \end{center}
 \end{figure}

In Fig. \ref{rho3} we compare the quasiparticle density of states at $n_d=1$
and $n_d=2$ for $U/\pi\Delta=2,4$, $J_{\rm H}=0.05$ and $\pi\Delta=0.01$.
It illustrates both the shift in the peak  the Kondo resonance to above
 the Fermi level for $n_d=1$,  the  enhancement of the density of states
at the Fermi level, and the narrowing of the resonance for the larger value of $U$.\par

\vspace*{0.7cm}

\begin{figure}[!htbp]
   \begin{center}
     \includegraphics[width=0.42\textwidth]{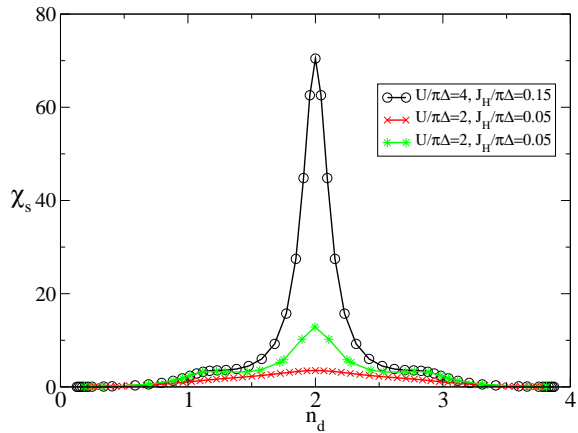}
     \caption{ (Color online) A comparison of the
spin susceptibility $\chi_s(\omega)$ (in units of $8\mu_{\rm B}^2$)
for the sets of renormalized parameters given in Figs. \ref{rdata1}, \ref{rdata2} and \ref{rdata3}.  } 
     \label{chi3}
   \end{center}
 \end{figure}

Finally in Fig. \ref{chi3} we compare the spin susceptibilities as a function of
$n_d$ calculated for the three sets of renormalized parameters given in Figs.
\ref{rdata1}, \ref{rdata2} and \ref{rdata3}.  The spin susceptibility is considerably enhanced
near particle-hole symmetry for the case $J_{\rm H}/\pi\Delta=0.15$
compared with that for $J_{\rm H}/\pi\Delta=0.05$ and the same value of $U$
($U/\pi\Delta=4$). It also illustrates the more modest enhancement of
$\chi_s$ for the larger value $U/\pi\Delta=4$ compared with the case for
$U/\pi\Delta=2$ and the same value of $J_{\rm H}$ ($J_{\rm H}/\pi\Delta=0.05$).
\vspace*{0.7cm}

\section{Conclusions}

The combination of the renormalized perturbation theory with explicit
calculations of the renormalized parameters from the numerical renormalization
group for $n=2$ have given us a comprehensive picture of the low
energy behavior of the $n$-channel Anderson model, with and without a Hund's
rule coupling term. One or two features of these results deserve some
additional discussion.

\par
One surprising feature revealed by the NRG calculations of the renormalized
parameters is the suppression
of the impurity charge fluctations over the whole range $1\le n_d \le 3$
for large $U$. At the points $n_d=1,2,3$ for large $U$, 
in the atomic limit the impurity levels are well away from the Fermi level,
and the quasiparticle resonance at the Fermi level is a many-body effect
induced by the spin fluctuations. In intermediate valent situation between
$n_d=1$ and $n_d=2$, and similarly between
 $n_d=2$ and $n_d=3$, in the atomic limit there are atomic excitation levels
 at the Fermi level, so once the hybridization is included one  might expect the electrons to jump on and off the
impurity
site relatively  freely giving a largely unrenormalized charge susceptibility.
The fact that  $\eta_c$ is very small in this range suggests that there
is a binding energy of electrons at the impurity site which suppresses
the local charge fluctuations, even in these intermediate valent regimes.
Presumably at temperatures much greater than this binding energy, the
mobility of the electrons in these intermediate valence regimes will
be restored. This topic deserves further investigation.\par

 Another feature of the model that deserves some comment is the behavior in the
large $n$ limit. We can contrast the situations for the SU(2n) model
when we are in the Kondo regime at half-filling when $n_d=n$,
with that for the model when $n_d=1$. This latter situation
corresponds to the infinite $U$ Anderson model used to describe
rare earth impurities. It can be seen from Eq. (\ref{relationa}) that in the limit
$n\to\infty$,  the effects of the interactions between the quasiparticles
goes to zero as $\tilde U\tilde\rho^{(0)}(0)\to 0$. The quasiparticle
interaction $\tilde U$ given by Eq. (\ref{uu}) remains finite in this
  limit.
The product $\tilde U\tilde\rho^{(0)}(0)$ tends to zero in this case,
where $n_d=1$, because
$\tilde\rho^{(0)}(0)\to 0$.  This contrasts with the situation
at half-filling where, as $n\to \infty$, the spin susceptibility $\chi_s$
must scale with $n$, so in taking the limit $n\to\infty$, $nT_{\rm K}$
must be kept constant. Hence, in this limit $\tilde\rho^{(0)}(0)$
remains finite and $\tilde U\to 0$. These different scenarios reflect
that the filling of the quasiparticle density of states must satisfy
the Friedel sum rule so that when $n_d=n$ it must span the Fermi level
symmetrically, while when $n_d=1$, the Fermi level must lie in the tail
of the quasiparticle density of states to give a filling factor
in each spin and channel of $1/2n$. These differences have physical
consequences for the low temperature  and magnetic field dependence  of the
susceptibility, giving a low energy peak when  the Fermi level
lies in the tail of the quasiparticle density of states, and also 
an enhanced thermopower.

\par

\bigskip
\noindent{\bf Acknowledgment}\par
\bigskip
{We thank Akira Oguri and Johannes Bauer for helpful discussions. Two of us
  (DJGC and ACH) thank the EPRSC for support (Grant No. EP/G032181/1)
and YN thanks the JSPS Grant-in-Aid for Scientific Research (C) support  (Grant No. 20540319).
The numerical calculations were partly carried out on SX8 at  YITP in Kyoto University.
\par
\bigskip

\bibliography{artikel}
\bibliographystyle{h-physrev3}

\end{document}